\documentclass{ws-procs9x6}
\def\etal {{\it et al.}}

\newcommand{\isotope}[2]{$^{#2}$#1}

\newcommand{\yr}{yr$^{-1}$}

\begin{document}
\title{TIME--VARYING NUCLEAR DECAY PARAMETERS AND DARK MATTER}
%%%%%%%%%%%%%% AUTHORS %%%%%%%%%%%%%%
%%%%%%%%%%%%%%%%%%%%%%%%%%%%%%%%%%%%%
%%%%%%%%%%%%%%%%%%%%%%%%%%%%%%%%%%%%%
\author{
J.\ NISTOR,$^a$ E.\ FISCHBACH,$^{a*}$ J.T.\ GRUENWALD,$^a$ D.\ JAVORSEK,$^b$\\ 
J.H.\ JENKINS,$^c$ and R.H.\ LEE$^d$
}
\address{
$^a$Purdue University, Physics, West Lafayette, IN 47906, USA\\
$^b$Air Force Test Center, Edwards Air Force Base, CA 93524, USA\\
$^c$Texas A\&M University, Nuclear Engineering, College Station, TX 77843, USA\\
$^d$United States Air Force Academy, Physics, Colorado Springs, CO 80920, USA\\
$^*$E-mail: ephraim@purdue.edu
}
%%%%%%%%%%%%%%%%%%%%%%%%%%%%%%%%%%%%%
%%%%%%%%%%%%%%%%%%%%%%%%%%%%%%%%%%%%%
%%%%%%%%%%%%%%%%%%%%%%%%%%%%%%%%%%%%%
%%%%%%%%%%%%%%%%%%%%%%%%%%%%%%%%%%%%%
\begin{abstract}
Recently published data suggest a possible solar influence on some nuclear decay rates, including evidence for an annual variation attributed to the varying Earth-Sun distance. Here, we consider the possibility that the annual signal seen by the DAMA collaboration, and interpreted by them as evidence for dark matter, may in fact be due to the radioactive contaminant \isotope{K}{40}, which is known to be present in their detector. We also consider the possibility that part of the DAMA signal may arise from relic big-bang neutrinos.
\end{abstract}
%%%%%%%%%%%%%%%%%%%%%%%%%%%%%%%%%%%%%
%%%%%%%%%%%%%%%%%%%%%%%%%%%%%%%%%%%%%
%%%%%%%%%%%%%%%%%%%%%%%%%%%%%%%%%%%%%
%%%%%%%%%%%%%%%%%%%%%%%%%%%%%%%%%%%%%
%
\bodymatter

\section{Introduction}\label{intro}

Nuclear decay measurements reported by independent groups in 1986\cite{ref1} and in 1998\cite{ref1a} exhibited an anomalous oscillatory behavior. Once known systematic causes were ruled out, the results suggested that the decay constants $\lambda$ in some radioactive isotopes might themselves be time dependent. If this were the case, then the usual decay law would assume the form $dN(t)/dt = -\lambda (t) N(t)$, where $\lambda$ and $T_{1/2} = \ln 2 / \lambda$ would be explicitly time dependent. Interestingly, a significant number of the decay-rate variations observed to date can be attributed in one way or another to the Sun: e.g., an annual signal presumed to arise from the annual variation of the Earth-Sun distance\cite{ref2}, $R$, a (10--15) yr$^{-1}$ variation associated with the Sun's rotation,\cite{ref3,ref3a} a Rieger periodicity,\cite{ref4} and evidence for an association between solar storms and the decay rate of \isotope{Mn}{54}.\cite{ref5} A summary of existing data suggesting a time dependence of $\lambda (t)$ is presented in Table 2 of Ref.\ \refcite{ref5b}.

Although some of the decay-rate variations in that table may be due in part to seasonal variations in the sensitivities of various detectors (i.e., arising from variations in temperature, pressure, and humidity), it seems unlikely that such variations could account for the fact that experiments at different locations, using a variety of isotopes and detector types, report similar effects. Moreover, seasonal effects would not be expected to account for periodicities in the (10--15) \yr range, nor for short-term changes in decay rates coincident with solar storms. In what follows we will thus assume that the observed annual periodic dependence of $\lambda (t)$ for some nuclear decays arises from the Sun through some as-yet unknown mechanism.

\section{Dark matter}\label{DM}

There are numerous reasons to believe in the existence of dark matter (DM), such as the observation of flat galactic rotation curves. A recent review of the literature can be found in Ref.\ \refcite{ref6}. Among the many ongoing DM searches, the DAMA collaboration has presented the strongest evidence to date for the presence of a DM signal.\cite{ref7} However, when their results are parametrized in terms of the DM particle mass and its cross section for scattering off nuclei, the region allowed by DAMA in the resulting exclusion plot is ruled out by other experiments.\footnote{Some models allowing for inelastic WIMP-nucleon scattering and spin dependence may still account for the DAMA signal.} This has raised the question of whether the observed DAMA signal could have an alternate explanation. 

Since DAMA utilizes a large NaI(Tl) detector, which is known to have potassium contamination, the possibility has been raised that the DAMA signal may be attributable to the decay of naturally occurring \isotope{K}{40} which has a half-life of $1.26\times 10^9$ yr. To understand the relevance of \isotope{K}{40}, we note that the DAMA signal is an annually varying count rate in their detector which may be consistent with the fact that the Earth's speed relative to the hypothesized DM halo reaches a maximum around June 2, when the Earth's velocity adds maximally to that of the Sun, and a minimum on December 2 when the Earth's velocity points the opposite direction. The resulting annual variation in the DM flux (known as the annual modulation signature) could thus explain the annual signal reported by DAMA.

In contrast, the \isotope{K}{40} hypothesis attributes the annual signal seen by DAMA in their NaI(Tl) crystals to X-ray photons emitted in the detector itself from \isotope{K}{40} decays. There are 3 dominant decay modes of \isotope{K}{40}, of which two are most relevant. The first is an electron capture (EC) mode to the 1461 keV excited state of \isotope{Ar}{40}, followed by the electromagnetic decay to the ground state of \isotope{Ar}{40}. The signal for this mode, which has a branching ratio of 10.55$\%$, is the 1461 keV gamma accompanied by a $\sim$3.2 keV X-ray photon arising from the electromagnetic cascade which results from filling the $K$-shell hole left by the captured electron. The second mode of interest is the EC decay of \isotope{K}{40} directly to the ground state, again with the emission of a 3.2 keV X-ray photon but without an accompanying 1461 keV de-excitation gamma.\cite{pradler} This mode, which is estimated to have a small branching ratio ($\sim$0.2\%), has not yet been seen directly but may be the most interesting (see below). As it turns out, the annually varying DAMA signal is detected for X-ray photons in the 2--4 keV range, with a peak intensity near 3 keV, which is just the energy expected from the cascade following EC in \isotope{K}{40}. Although this may be coincidental, it may also be a clue to why the DAMA signal is not seen in other DM experiments. 

Another clue comes from the count rate observed by DAMA. Given the concentration of potassium that DAMA acknowledges to be present in their NaI(Tl) detectors, we estimate that \isotope{K}{40} would produce an event rate in the range of $\sim$1000 counts per day (cpd) from EC decay to the excited \isotope{Ar}{40} state, whereas for DAMA/LIBRA the total count rate in the 2--4 keV region is $\sim$530 cpd. Thus \isotope{K}{40} could correctly account for both the energy range in which DAMA see their signal and their observed count rate.

The preceding picture is suggestive, but far from complete, for the following reasons. (1) First, it presumes that \isotope{K}{40} exhibits an annual modulation in the EC decay mode, similar to other isotopes, for which there is no evidence at present. Fortunately, an experiment under way by Lang and collaborators should settle this question in the next year or so. (2) Secondly, the phase of the DAMA signal (i.e., the calendar date where their count rate is maximum) is approximately mid-May, whereas the isotopes exhibiting annual variations which have been studied to date have phases that are typically in the January-February or July-August time frames. This could be a problem for the \isotope{K}{40} hypothesis or, as we discuss below, a hint at other new physics. (3) Another problem  is the claim by DAMA that they see no annual modulation in the signal for those events in which both a 1461 keV gamma and a 3.2 keV X-ray (the expected signal for the EC decay mode) are detected.\cite{DAMA} Since the data to back up this claim have not been made available by the DAMA collaboration, their claim cannot be further evaluated at this time. Irrespective of this, it is worth noting that the modulation in the low-energy events accompanied by \isotope{K}{40} decay is {\em not} necessarily correlated with a modulation in the 1461 keV event rate. Those events which are associated with the EC decay directly to the ground state of \isotope{Ar}{40} (on the order of $\sim$20 cpd) are not associated with the emission of higher energy photons --- and therefore represent an {\em irreducible background} in the energy region in which DAMA claims a signal. The activity from this decay may be large enough to account for DAMA's observed annual modulation in count rate (amplitude of $\sim$10 cpd).

\section{Discussion}

Although the \isotope{K}{40} hypothesis as a mechanism to explain the DAMA data remains open, there may be a hint in the phase of the DAMA data of even more new physics. Consider a system sensitive to several influences with a common frequency $\omega$, but different amplitudes and phases, described by a function $f(t)$:
\begin{equation}\label{eq:f}
f(t) = f_{\mathrm{max}} \left[ 1 + \sum_i \delta_i \cos (\omega t - \phi_i ) \right] \equiv f_{\mathrm{max}} \left[ 1 + \alpha \cos (\omega t - \beta ) \right].
\end{equation}

In Eq.\ \eqref{eq:f}, $\tan \beta = (\sum \delta_i \sin \phi_i) / (\sum \delta_i \cos \phi_i) \equiv B/A$, and $\alpha = A / \cos \beta = B / \sin \beta$, and $\delta_i \ll 1$ is assumed. It is thus possible that the `anomalous' DAMA phase is the result of detecting several annual signals. These could include a solar influence (phase Jan-Feb or July-Aug), dark matter (May-Jun), and big bang relic neutrinos ($\sim$Dec 10). The possibility that DAMA could also be detecting relic neutrinos follows from the similarity of the fluxes of solar and relic neutrinos, as noted in Refs.\ \refcite{ref5} and \refcite{par}, notwithstanding their differences in energy.

\end{document}